\documentclass[prl,aps,preprint,showpacs,showkeys]{revtex4}
\begin{document}
\date{\today}
\title{Percolation in a Multifractal}
\author{  G. Corso, J. E. Freitas, L. S. Lucena, R. F. Soares}

\affiliation{ International Center for Complex Systems and
 Departamento de F{\'\i}sica Te\'orica e Experimental,
 Universidade Federal do Rio Grande do Norte, Campus Universit\'ario
 59078 970, Natal, RN, Brazil.}

\begin{abstract}

We investigate percolation phenomena in multifractal objects
that are built in a simple way. In these objects
the multifractality comes directly from the geometrical tiling.
 We identify some differences between percolation in the proposed
 multifractals  and  in a regular lattice.
 There are basically two sources of these differences.
 The first is related with the
coordination number that
 changes along the multifractal. The second comes from the way
 the weight of each cell in the multifractal
affects the percolation cluster.
We use  many samples of finite size lattices and draw
the histogram of percolating lattices against site
occupation probability.  Depending on a parameter
 characterizing the multifractal and the lattice size,
 the histogram can have two peaks. We observe that the percolation threshold for the
multifractal is lower than one for the square lattice.
   We compute the
fractal dimension of the percolating cluster and the  critical exponent $\beta$.
Despite the topological differences, we find that the percolation in a
multifractal support is in the same
 universality class of the standard percolation.

\end{abstract}

\pacs{ 64.60.Ak   61.43.Hv   05.40.-a  91.60.-x}

\keywords{ percolation,  multifractal,  complex systems, universality class}

\maketitle

\section{1- Introduction}
\hspace{\parindent}
Percolation theory has been used in several fields such as
chemistry, epidemics, science of materials,
transport of fluids in porous media, branched polymers,
and econophysics \cite{perc1,stauffer,stanley,perc3,bp1,bp2,perc4}. 
The original percolation model based on a square lattice has 
been extended to several kinds of regular and random lattices,  to 
continuous media where the objects  overlap in space, and to  
other complex systems \cite{complex1,complex2,sahimi1,sahimi2}. 
In this work  we generalize the percolation theory to cover an even broader
range of complex systems.  We devised an approach to
investigate how percolation occurs in a support that is itself a multifractal.  
For this purpose we have constructed an easy assembling
multifractal immersed in a $2D$ space.

Our work is inspired in the modeling of geophysical natural objects that 
show multifractal properties    
 \cite{Muller1,Muller2,Riedi,Hubert}. 
The model can  be applied to
 transport of fluid in a multifractal porous media such as  
 sedimentary strata. 
Oil reservoirs are possible candidates to be modeled in such a way
 since that the measurement of some physical quantities in well-logs 
show multifractal behavior \cite{Lovejoy,Hermann}.   
Despite the potential applications inspiration,
this problem is important by itself in the scientific context.
 The study of
percolation phenomena in multifractal lattices is relevant in Statistical 
Physics specially when the size of the blocks and their 
number of neighbors can vary. 

 In order to make this analysis we  create a
multifractal object that can be used as a toy model and 
 a laboratory for  percolation theory. 
An important characteristic of this object is that
its topologic properties
(e.g. number of neighbors of each block) change over the object.
In the reference \cite{sahimi3} an algorithm that has some
resemblance with ours is used. That multifractal is built
from the partition of the square,
but the object has a trivial topology. Besides, the object
used in \cite{sahimi3} is aleatory and ours is deterministic.
Although both models present multifractality, our model has the
following differences: it shows a non-trivial topology, we can determine
analytically its spectrum of fractal dimensions,
it generalizes the square lattice,
 and it shows  simplicity in the construction.

The multifractal object we have developed is
a natural generalization of the regular square lattice
once we consider the algorithmic point of view.
 The algorithm that generates a square lattice with
$2^n \> \times 2^n$ cells  starting with a square of
fixed size is the following.
We begin with a $L \times L$ square and we cut if in $4$
identical pieces (cells).
 At each step all the cells are equally divided in $4$ parts using
  vertical and  horizontal segments. This
 process produces a lattice as a partition of the square.
 The multifractal we create is also a partition of
the square, but the ratio we
divide the cells is different from $L/2$. The parameter
characterizing the multifractal,
$\rho$, is related to the ratio of this division.

What makes this problem new and appealing to physics is
the following. The support of the percolation
clusters is composed by subsets of different fractal dimensions.
It is important to know how these different subsets are connected and
how they participate in the conducting process. There are intriguing
features in the network due to the fact that all cells have
rectangular shape but the area and the number of neighbors can vary,
forming an exotic tilling.

 In section $2$ we present the multifractal object that we construct
 to study percolation, and we analyze how its multifractal partition maps
   into the
square lattice. In section $3$ we expose the algorithm we use to estimate  
percolation threshold and derive the multifractal spectrum of the multifractal 
object.
In section $4$ we show numerical results and discuss the histograms of percolating
lattices versus occupation probability. Finally in
section $5$ we summarize the main differences
between percolation in a regular lattice and in a multifractal support.

\section{2- The multifractal object $ Q_{mf}$}
\hspace{\parindent}
The central object of our analysis is a multifractal object that
we call $ Q_{mf}$. Before defining it we enumerate some of its properties.
\begin{itemize}
\item $ Q_{mf}$ is a multifractal, it means,  $Q_{mf}$ has an infinite number of
 $k$-subsets  each one with a distinct fractal dimension $D_k$.

 \item It is possible to  determine analytically the spectrum of all $D_k$.

\item  The sum of all the families of
 $k$-subsets fills the square. This fact enables us
 to study its percolation properties using procedures similar to
the ones applied in the site percolation in the square lattice.

\item The algorithm of construction of $ Q_{mf}$ has just one parameter $\rho$.

\item For the special choice  $\rho=1$ the object $ Q_{mf}$ degenerates into the
square lattice. In this case we compare 
 our results with the square lattice site percolation.

\item The object $Q_{mf}$ shows self-affinity or self-scaling depending on
the region of the object.

\item  Finally, the algorithm for construction of $ Q_{mf}$ is
simple and it is of easy implementation in the computer.

\end{itemize}

We define $ Q_{mf}$ through the following algorithm.
 We start with a square of linear size $L$ and choose a parameter
$0<\rho<1$, where $\rho=\frac{s}{r}$ for $r$ and $s$ integers. 
In the first step, $n=1$,
the square is cut in two pieces of area $\frac{r}{s+r} = \frac{1}{1+\rho}$ 
and $\frac{s}{s+r} = \frac{\rho}{1+\rho}$ by a
vertical line (we use $L^2$ units). 
In other words, the square is cut
according to a given  $\rho$.
This step is shown in figure \ref{fig1} (a), in this figure we use
$\rho = \frac{s}{r} = \frac{2}{3}$.

In the second step, $n=2$, we cut  the two rectangles of
figure \ref{fig1} (a) by the same  $\rho$, but
using two horizontal lines as shown in figure \ref{fig1} (b).
This partition of the square generates
four rectangular blocks: the smallest one of area $(\frac{\rho}{1+\rho})^2$,
two of them of area $\frac{\rho}{(1+\rho)^2}$ and the
largest one of area $(\frac{1}{1+\rho})^2$, in the
figure $\rho >0.5$.

The third step, $n=3$, is shown in figure \ref{fig1} (c) and
the fourth step, $n=4$, in (d).
 As observed in  figure,
at level $n=4$ there are $2^4$ blocks
and the distribution of areas among the blocks follows the
binomial law:
\begin{equation}
         1= (\frac{\rho}{1+\rho})^4 +4(\frac{\rho}{1+\rho})^3 (\frac{1}{1+\rho})
+6 (\frac{\rho}{1+\rho})^2 (\frac{\rho}{1+\rho})^2 
+4 (\frac{\rho}{1+\rho}) (\frac{1}{1+\rho})^3 +(\frac{1}{1+\rho})^4.
\label{bino1}
\end{equation}
We call the elements with the same area as a $k$-set. In the case $n=4$ we have five $k$-sets.

 At step $n$ the square
  has $2^{n-1}$ line segments, $(n+1)$ $k$-sets and
$2^n$ blocks. The partition of the area $A=1$ (using $L^2$ units)
 of the square in different blocks
 follows the binomial rule:
\begin{equation}
         A= \sum_{k=0}^n C_k^n (\frac{\rho}{1+\rho})^k (\frac{1}{1+\rho})^{n-k} 
= (\frac{1+\rho}{1+\rho})^n = 1.
\label{bino2}
\end{equation}

As $n \rightarrow \infty$ each $k$-set
(a subset made of cells of same area)
determines a monofractal whose dimension we
calculate in the next section. The ensemble of all
$k$-sets engenders the multifractal object $Q_{mf}$.

\section{3- The algorithm of percolation and the multifractal spectrum}
\hspace{\parindent}
In this section we show the algorithm used to study the percolation properties
of $ Q_{mf}$ and the analytic derivation of its spectrum of fractal dimensions.
The estimation of the spectrum, $D_k$, is performed using the box counting method
\cite{barnsley} whose measure elements came from the percolation algorithm.

The concept of the percolation algorithm for  $Q_{mf}$
consists in mapping it into the
square lattice. The square lattice should be large enough that
each line segment of $Q_{mf}$
coincides with a line of the lattice.
Therefore we consider that the square lattice is more
finely divided than $Q_{mf}$. In this way all blocks
 of the multifractal are
composed by a finite number of cells of the square lattice.

To explain the percolation algorithm we suppose that  $Q_{mf}$
construction is at step $n$.
We proceed the percolation algorithm by choosing at
random one among the $2^n$ blocks of
$ Q_{mf}$. Once a block is chosen all the  cells
in the square lattice corresponding to this block are considered as
occupied. Each time a block of $ Q_{mf}$ is chosen
the algorithm check if the occupied cells at the underlying
lattice are connected in such a way to form an $infinite$ percolation
cluster.
The algorithm to check the percolation is similar to the one used in
 \cite{ziff,freitas0,freitas1,freitas2}.

For the estimation of  the spectrum $D_k$ of an object $X$ we use the box
counting method \cite{barnsley}. The object $X$ is
immersed in the plane of real numbers
 $\Re^2$, we use the trivial metric. Cover $\Re^2$
  by just-touching square boxes
of side length $\epsilon$. Let $N(X)$ denote the number of square cells
 of side length $\epsilon$
which intersect $X$. If
\begin{equation}
                 D_X = lim_{\epsilon \rightarrow 0}
          \frac{log \> N(X)}{log \> \frac{1}{\epsilon}} 
            = lim_{L \rightarrow \infty} \frac{log \> N(X)}{log \> L }  ,
\label{boxcou1}
\end{equation}
 is finite, then $D_X$ is the dimension of the $X$.

 In our case the object $X$ is a $k$-set. Remember that
the $k$-set corresponds to a set of rectangles of
the same area. For a $k$-set we have that $N_k$ is given by:
\begin{equation}
                 N_k = C_n^k \> \> s^k \>r^{(n-k)},
\label{NNN}
\end{equation}
where $C_n^k$ is the binomial coefficient that express
the number of elements $k$-type, and
$s^k \> r^{(n-k)}$ is the area of each element of this set. If the square is
partitioned $n$ times ($\frac{n}{2}$ horizontal cuts and 
$\frac{n}{2}$ vertical cuts) its size is $L=(s+r)^{n/2}$.
Combining all this information we have for the fractal dimension of
each $k$-set:
\begin{equation}
                 D_k = lim_{n \rightarrow \infty} \frac{log \> C_n^k \>
\> s^k \>r^{(n-k)} }{log \> (s+r)^\frac{n}{2} },
\label{boxcou2}
\end{equation}

In the $r=s=1$ case all subsets of $ Q_{mf}$ are composed by elements of the same area,
square cells. In this way the object is formed by a single subset with dimension:
\begin{equation}
                 D = lim_{n \rightarrow \infty} \frac{log \>
\> (1+1)^n }{log \> (1+1)^\frac{n}{2} } = 2,
\label{boxcou76}
\end{equation}
This result is expected once in this particular case $ Q_{mf}$ degenerates
into the square lattice that has dimension $2$.

In figure \ref{fig2} we show the picture of
$ Q_{mf}$ for $\rho = \frac{2}{3}$. We have used
$n=12$, in (a) the full object is shown, in (b) a zoom of an
internal square of the object is illustrated.
 We have used the same
color to indicate the elements of a same $k$-set.
The funny tilling depicted in the
figure is common to $ Q_{mf}$ with different values of $\rho$.

Figure \ref{fig3} shows the spectrum of $D_k$ for $n=400$
calculated from equation (\ref{boxcou2}). The use of increasing $n$
does not change the
shape of the curve, it only increases the number of $k$ and makes
the curve appear more dense. We use
$(s,r)=(2,3)$ to illustrate the asymmetry of the distribution.
The spectrum has a maximum close to
$\rho \> n$.
In this case $\frac{2}{3} 400 \simeq 270$.
It means that the majority of mass of the multifractal is
concentrated in the $k$-sets around this value.
The spectrum $D_k$ is typically asymmetric around
its maximum. Only the case
$(s,r)=(1,1)$ is symmetric and the asymmetry of $D_k$ increases
as $\frac{s}{s+r} \rightarrow 1$, which is
related to the area distribution among the blocks as we shall see in the next chapter.

\section{4- Numerical Simulations}
\hspace{\parindent}
In this chapter we focus our attention on the numerical
results obtained from the algorithm
exposed above. We are interested mainly in analyzing the percolating
properties of $Q_{mf}$. Figure \ref{fig4} (a) shows the
histogram of percolating lattices versus the occupation probability $p$.
The area under the histogram is normalized to unity. We use $n=10$ and
average the results over $40000$ samples. We consider that a lattice
 percolates when it percolates from top to
bottom or from left to right. The histogram of percolating lattices 
 at both directions
is similar but slightly shifted to the right. This shift is
common in percolation (see the reference  \cite{ziff} for
percolation in the square lattice).

We show in figure \ref{fig4} (a) the results of simulations for
 the following values of $(s,r)$: $(1,1)$,
which degenerates into the square lattice;
and $(2,1)$, $(4,1)$, $(6,1)$ which correspond to truly
multifractals. In this figure the histograms  corresponding to
 $(2,1)$, $(4,1)$ and $(6,1)$
are shifted to the left compared to the histogram of  $(1,1)$.
 The peak of the histogram for $(1,1)$ corresponds, as expected,
to the square lattice size percolation threshold, $p_c=0.597$, \cite{stauffer},
since this case matches exactly the square lattice.
The other values of $p_c$ are shown in Table $I$.

  The reason why $Q_{mf}$, for diverse $\rho$, shows roughly the same $p_c$
comes from the topology of the multifractal. The topology of
a set of blocks is related with
the coordination number, $c$, which is defined as the number
of neighbors of each block \cite{stauffer}.  $Q_{mf}$ has the property that
 $c$ changes along the object and with $\rho$. However, we
compute the average coordination number $c_{ave}$.
These results  do not depend   significantly
 on $\rho$, neither on  $n$,  the number of steps to build $Q_{mf}$,
which determines the number of blocks.
The value found, ${\it c}_{ave}=5.436$, for the multifractal
is close to the value of $c$ of the triangular
 percolation problem which has $c=6$ and whose analytic percolation
threshold is $p_c=0.5$. 
 The situation $(s,r)=(1,1)$, the square lattice, shows trivially
$c=4$. Because the square lattice has
a different $c$ it configures a particular situation compared to
other $Q_{mf}$ and it shows a different $p_c$ as depicted in figure
\ref{fig4} (a).

In table $I$ we show $p_c$ and the fractal dimension of the percolating
cluster, $d_f$, for diverse $\rho$. We have done an average over $100000$
samples and $n=16$. The estimation of $d_f$ is done by the relation
$M \sim L^{d_f}$ for  the \lq mass \rq\ ,$M$ ,  of the percolation cluster, that
means, the area of the cluster measured in units of the underlying square
lattice, and $L$, the
size of the underlying lattice. Based on the values of $d_f$ of
table $I$ we conclude that the
percolation on a multifractal support (imbeded in two dimensions)
 belong to the same class of
universality than the usual percolation in two-dimensions. The calculated value of $d_f$ for the
$(6,1)$ case is smaller compared to the others because of finite size effects. We
discuss in detail this effect in the following paragraphs.

Percolation shows  critical phenomena and several
scaling relations are observed.
The critical exponent $\beta$  is defined
from the equation:

\begin{equation}
                 R_L \sim \left( p_c(L) - p_c \right)^{\beta},
\label{beta}
\end{equation}
where $p_c$ is the exact occupation probability value in contrast to
 $p_c(L)$ which is the finite size value. The power-law (\ref{beta})
 is verified for $p_c(L)$ obtained  from $R_L$.
The numerical estimation of $\beta$ is based in equation (\ref{beta}) where
$R_L$ is a key element of the analysis. For $Q_{mf}$ the probability
 $R_L$  is not a well behaved function of $p$ for low $L$ as we shall see in
the next paragraphs.
Actually, $R_L$  can show, depending on $\rho$, an inflection
point at $p_c$ in this regime. However, in the
case where $L \rightarrow \infty$ the scaling of $\left( p_c(L) -p_c \right)$
recovers an usual behavior. In this regime we find the same $\beta$
characteristic of two dimension case,
$\beta=\frac{5}{36}=0.13888$. We checked in our simulations that for $n=18$,
 $\beta$ is around $5\%$ of the exact
value. The full set of values of $\beta$ is in table $I$.

{\it Table I}
\begin{tabbing}
  $(s,r)$      \hspace{0.6cm}        \= (1,1) \hspace{0.4cm} \= (2,1) \hspace{0.4cm}\=      (3,1) \hspace{0.4cm} \= (3,2) \hspace{0.4cm} \= (4,1) \hspace{0.4cm} \=(5,1)  \hspace{0.4cm} \= (6,1) \\
\hspace{0.3cm} $p_{c}$  \hspace{0.7cm} \= 0.593 \hspace{0.2cm} \= 0.527  \hspace{0.3cm} \= 0.526  \hspace{0.2cm} \= 0.526   \hspace{0.2cm} \=  0.525 \hspace{0.3cm} \= 0.525  \hspace{0.3cm} \= 0.530  \\
\hspace{0.3cm} $ d_f$   \hspace{0.7cm} \= 1.895 \hspace{0.2cm} \= 1.900 \hspace{0.3cm} \= 1.911   \hspace{0.2cm}  \= 1.890  \hspace{0.2cm} \= 1.902  \hspace{0.3cm} \= 1.929  \hspace{0.3cm} \= 1.842  \\
\hspace{0.3cm} $ \beta$   \hspace{0.8cm} \= 0.127 \hspace{0.2cm} \= 0.128 \hspace{0.3cm} \= 0.140   \hspace{0.2cm}  \= 0.141  \hspace{0.2cm} \= 0.141  \hspace{0.3cm} \= 0.118  \hspace{0.3cm} \= 0.109  \\

\end{tabbing}

It is worth to say that, despite small fluctuations in the values
shown in the table, there is no tendency in the numbers. The conclusion we
take from this data is that the errors are caused by finite size effects
and low-number statistics.

The dispersion of the
histogram changes significatively with $(s,r)$ as intuitively expected.
To illustrate the changing in the width of the histogram of a generic
$(s,r)$ multifractal we analyze the area of its blocks. At step $n$ of the
construction of  $Q_{mf}$ the largest element has the area
$ \frac{s^n}{(s+r)^n}$ and the smallest 
$ \frac{r^n}{(s+r)^n}$ (using $L^2$ units). 
In this way the largest area ratio among blocks increases with
$(\frac{s}{r})^n$. As the
occupation probability, entering in
the percolation algorithm, is in general proportional to the
area of the blocks we expect that the width of the
histograms in figure \ref{fig4} (a)
increases with $(\frac{s}{r})^n$. This increase in the
dispersion is visualized clearly
in the curves $(2,1)$ and $(4,1)$ of the figure.

The most singular curve in figure \ref{fig4} (a) is $(6,1)$ which
shows clearly two peaks. We stress this point when we comment
figure \ref{fig5}.
Figure \ref{fig4} (b) uses the same data of figure (a), but instead of
the histogram of percolating lattices we show the
cumulative sum,
$R_L$. As $R_L$ is normalized, this parameter is also called
the fraction of percolating lattices.
 As in \ref{fig4} (a) the case $(s,r)=(1,1)$, the square lattice,
reproduces the results of literature \cite{ziff}. In this situation the lattice
  size, $L$, is $L=(s+r)^{10}=1024$. For this special
  case the number of blocks is equal to the number of unit boxes covering the
 surface. The double peak case
  $(s,r)=(6,1)$ shows an inflection point in the graphic of $R_L$ versus
  $p$. In the following figure we explore in detail this point.

The most  noticeable signature of percolation in the multifractal $Q_{mf}$ is
 the double peak observed for $(s,r)=(6,1)$ in figure
\ref{fig5}. In this figure the histograms of percolating lattices versus
$p$ is plotted for diverse $n$ as indicated in the figure. The distance between
the peaks decreases as $n$ increases. This picture indicates
that the double peak is a phenomenon that
is relevant for percolation in the multifractal, when 
$\rho$ is low, in
the finite lattice size condition
used in the simulation.
From an analytic point of view the curve $(6,1)$ in figure \ref{fig5} is different
from curve $(1,1)$. In curve $(6,1)$ there are three extrema points while in
the $(1,1)$ case the curve shows a single maximum point. We conjecture that in the
limit of $n \rightarrow \infty$ these three points coalesce into
 a single one and all
the curves show a similar behavior.

The two peaks in the histogram come from the huge difference among the area of
the blocks of $Q_{mf}$. For large $(\frac{s}{r})^n$ the area
difference is so accentuated
that we model the histogram of percolating
lattices as a bimodal statistics. In the
case of the largest block be chosen the multifractal easily percolates compared
with the opposite possibility. To estimate the effect of the largest area block
on the statistic we make Table $II$. The
difference between the first peak at $p_1$ and
the second one at $p_2$ is $\Delta p_{max}$. In Table $II$ we
compare  $\Delta p_{max}$
with the fraction of the largest block over the total
square area $( \frac{s}{s+r})^n$.
This comparison is made for different steps in the construction of the
multifractal $n$, as $n$ increases the area difference
decreases as well as the distance
between peaks.
Table $II$ shows a good agreement between both values, we conclude that the
bimodal statistic is  caused by the huge mass of the largest block.

{\it Table II}
\begin{tabbing}
 $n$  \hspace{1.4cm} \= 8 \= \hspace{1.5cm} 10 \= \hspace{1.5cm} 12 \= \hspace{1.5cm} 14  \= \hspace{1.5cm} 16 \= \hspace{1.5cm} 18 \\
 $\Delta p_{max}$  \hspace{0.4cm} \= $0.29$ \= \hspace{1.1cm} $0.22$ \= \hspace{1.1cm} $0.15$   \=  \hspace{1.1cm} $0.11$ \= \hspace{1.1cm} $0.070$ \= \hspace{1.1cm} $0.040$\\
 $( \frac{s}{s+r})^n$   \hspace{0.4cm} \= $0.291$ \= \hspace{0.9cm} $0.211$ \= \hspace{0.9cm} $0.157$   \=  \hspace{0.9cm} $0.115$ \= \hspace{0.9cm} $0.084$ \= \hspace{1.1cm} $0.062$\\
 \end{tabbing}

 We notice, however, that the agreement between
$\Delta p_{max}$ and $(\frac{s}{s+r})^n$
 decreases as $n$ increases. We interpret the disagreement
between the bimodal statistics hypothesis
 and the numerics for high $n$ as the limit of the hypothesis. Actually, the
 largest block is not the only one that produces
anisotropy in the multifractal, and as
 $n$ increases this fact becomes more
 accentuated. For small $n$ the large block can be
taken as the main
 factor of anisotropy, and the bimodal statistics
apply. Large $n$ implies, however,
 truly multifractals and a more complex statistics
should be used to treat the problem.

\section{5- Conclusion}
\hspace{\parindent}
   In this work we develop a multifractal object, $Q_{mf}$
to study percolation
in a multifractal support. Besides of $Q_{mf}$ being a multifractal,
it shows several interesting properties. The sum of all its
fractal subsets fills a square and it is possible
to determine the spectrum of its fractal dimensions.
In addition, the algorithm that
generates $Q_{mf}$ has only one free parameter $\rho$, and in
the  $\rho=1$ case $Q_{mf}$ becomes the square lattice.

We observe that percolation in a multifractal presents
different features from
percolation in a regular lattice.
There are two reasons for that:
the heterogeneous distribution of weight (area)
among the blocks and
the variation of the coordination number of the topologic structure.
  The weight of each block in a multifractal
 counts diversely in the mass of the infinite
  percolating cluster. The difference
 in weight of the blocks changes the dispersion of the histogram of
 percolating lattices. The phenomenon of
 two peaks appearing in the histogram is also connected with the
 weight difference. We model the distance
between peaks using a bimodal statistics.
 In the limit of $n \rightarrow \infty$
 all the histograms of multifractals seems to collapse
into a single curve.

For all cases in which $\rho \neq 1$
the multifractal $Q_{mf}$ shows a coordination number
(number of neighbors of each block)
that changes along the object. The average coordination number of
$Q_{mf}$ is around $5.436$.
In contrast, the situation $(s,r)=(1,1)$ (the square lattice particular case)
has a coordination number constant and equal to $4$.
This suggests that the case $\rho \neq 1 $ represents a break of
the symmetry of the system.
In this sense the coordination number (topology)
is much more complex for $Q_{mf}$ than for a regular lattice.
Despite these differences, we have done numerical estimations
of the fractal dimension of
the percolating cluster in the multifractal,
obtaining values which are around $1.89$,
the same dimension found for the incipient percolation
cluster in  a two-dimensional regular lattice. The numerical simulation
of the $\beta$ critical exponent, also show the same value of
two-dimensional regular case, and point to the same conclusion that
we have regular percolation.


\vspace{2cm}

The authors gratefully acknowledge the financial support of Conselho Nacional
de Desenvolvimento Cient{\'\i}fico e Tecnol{\'o}gico (CNPq)-Brazil,
FINEP and CTPETRO. We also thanks to D. M. Tavares for the helpeful 
comments.

\centerline{FIGURE LEGENDS}

\begin{figure}[ht]
\begin{center}
\caption{ The four initial steps in the formation of $Q_{mf}$. In (a) the vertical line cutting the
square in two pieces of area ratio $\rho$. Two horizontal lines sharing the rectangles by
 the same ratio are depicted in (b). The third  step is indicated in (c)
and the fourth in (d). At each
 step the areas of the respective blocks are shown in figure.}
\label{fig1}
\end{center}
\end{figure}
\begin{figure}[ht]
\begin{center}
\caption{The figure shows the three views of the multifractal $Q_{mf}$,
it is used  $n=12$, $(s,r)=(3,2)$. In (a) we have the original picture,
figure (b) is a zoom of the square indicated in (a).  }
\label{fig2}
\end{center}
\end{figure}

\begin{figure}[ht]
\begin{center}
\caption{The spectrum of fractal dimensions $D_k$ of $ Q_{mf}$ for
$n=400$ and $(p,q)=(3,2)$.}
\label{fig3}
\end{center}
\end{figure}
\begin{figure}[ht]
\begin{center}
\caption{In (a) is depicted the histogram of percolation lattices versus the
 occupation probability $p$ for
 the cases $(s,r)=(1,1)$, $(2,1)$, $(4,1)$, and $(6,1)$.
 The areas under the curve are normalized to
unity. For the same (s,r) is shown in (b) the graphic of
the fraction of percolation lattices $R_L$ versus $p$. It is used $40000$ lattices to
make the average.}
\label{fig4}
\end{center}
\end{figure}
\begin{figure}[ht]
\begin{center}
\caption{The histogram of percolated lattices
versus the occupation probability $p$
for several values lattice size. The graphic
shows the double peak approaching each other as
$n$ increases, in the figure $(s,r)=(1,6)$ and $8<n<18$.
It is used $40000$ lattices to make the average.}
\label{fig5}
\end{center}
\end{figure}

\end{document}